\documentclass[12pt]{article}

\usepackage[utf8]{inputenc}
\usepackage[T1]{fontenc}
\usepackage{authblk} 
\usepackage{amsmath,amsfonts,amssymb,amsthm}
\usepackage{dsfont} 
\usepackage{color}
\usepackage[dvipsnames]{xcolor} 
\colorlet{mygreen}{green!60!black}
\colorlet{mygrey}{white!60!black}
\usepackage{graphicx} 
\usepackage{float}  
\usepackage{booktabs} 
\usepackage{setspace} 
\usepackage{multicol} 
\usepackage{url} 
\usepackage{hyperref}
\usepackage{calc} 

\usepackage{wrapfig}
\usepackage{lscape}
\usepackage{rotating}
\usepackage{epstopdf}

\usepackage{tikz}
\usetikzlibrary{shapes,positioning} 
\usetikzlibrary{calc} 
\usetikzlibrary {arrows.meta} 

\newlength{\wirespace} 
\newlength{\wirespaceh} 
\newlength{\wirespaceht} 
\newlength{\wirespacehb} 
\newlength{\boxsize}
\newlength{\messpace} 
\usepackage{geometry} 
\usepackage[font=small,labelfont=bf]{caption} 
\usepackage{palatino} 

\newcounter{quotecounter}

\newcommand{\lquote}[3]{%
    \refstepcounter{quotecounter} 
    \par\noindent
    \begin{center} 
    \hspace{35pt}
        \begin{minipage}{0.8\textwidth} 
            \textit{#3} 
        \end{minipage}%
        \hfill 
        \begin{minipage}{0.1\textwidth} 
            \begin{flushright}
                \text{(#2)}
                \label{#1} 
            \end{flushright}
        \end{minipage}
    \end{center}
}

\newcommand{\qref}[1]{\text{(Q\ref{#1})}}

\theoremstyle{definition}

\theoremstyle{theorem}

\def \be {\begin{equation}} 
\def \ee {\end{equation}}
\def \bes {\begin{equation*}}
\def \ees {\end{equation*}}
\def \baa {\begin{align}}
\def \eaa {\end{align}}
\def \baas {\begin{align*}}
\def \eaas {\end{align*}}
\def \bea {\begin{eqnarray}}
\def \eea {\end{eqnarray}}
\def \beas {\begin{eqnarray*}}
\def \eeas {\end{eqnarray*}}

\newcommand{\ket}[1]{\rvert#1\rangle}

\newcommand{\ketbra}[2]{| #1 \rangle \langle #2 |}

\newcommand{\one}{\mathds{1}}

\newcommand{\bol}{\boldsymbol} 

\title{On the Alleged Locality\\
 in the Schrödinger Picture
}
\author{Charles Alexandre B\'edard}
\affil{\small {École de technologie supérieure}\\
\footnotesize \emph{charles.alexandre.bedard@etsmtl.ca}}
\date{\today\vspace{-10pt}}

\begin{document}
\maketitle
\thispagestyle{empty}
\begin{abstract}
\noindent  
Vedral claims~\cite{vedral2024locality} that the Schrödinger picture can describe quantum systems as locally as the Heisenberg picture, relying on a product notation for the density matrix. Here, I refute that claim. I show that the so-called `local factors' in the product notation do not correspond to individual systems and therefore fail to satisfy Einsteinian locality.
Furthermore, the product notation does not track where local gates are applied. Finally, I expose internal inconsistencies in the argument: if, as is also stated, Schrödinger-picture locality ultimately depends on explicit bookkeeping of all operations, then the explanatory power of the product notation is \emph{de facto} undermined.
\end{abstract}

\section{Motivation}\label{secmotivation}


In the Heisenberg picture, quantum systems can be fully described through local descriptions of their parts, even when the parts are entangled 
\cite{gottesman1998heisenberg, deutsch2000information, bedard2021abc}.
A bipartite system~$AB$ admits \emph{descriptors}~$\bol q_A$ and~$\bol q_B$ that are
\begin{itemize}
\item[(a)] \emph{Einstein-local}:~$\bol q_A$ is independent of what is done to system~$B$, and \emph{vice versa}, and;
\item[(b)] \emph{empirically complete}:~$\bol q_A$ and~$\bol q_B$ encompass sufficient information to calculate the probability distributions associated with any measurement performed on the joint system~$AB$.
\end{itemize}

In the Schrödinger picture, while reduced density matrices~$\rho_A$ and~$\rho_B$ fulfill property~(a) above, they fail to satisfy property~(b): the probability distributions of measurement outcomes on~$AB$, encompassed in~$\rho_{AB}$, cannot be obtained from the local~$\rho_A$ and~$\rho_B$ when the system is entangled.
On the other hand, should one adopt~$\rho_{AB}$ as a description of each individual system, empirical completeness would be trivially obtained at the price of losing Einsteinian locality.
Thus, the Schrödinger picture appears to impose a dichotomy: a system can be described either locally or completely, but not both.
As mentioned, the Heisenberg picture resolves this dichotomy with descriptors that satisfy both (a) and (b).

In \emph{Locality in the Schrödinger Picture of Quantum Mechanics}~\cite{vedral2024locality} Vedral suggests that the Schrödinger picture can also circumvent the dichotomy. Upon writing the global density matrix in a so-called product notation, Vedral claims that 
\lquote{localfactors}{Q1}{
[T]he state in the product notation is as local in the Schrödinger picture as it is in the Heisenberg picture, meaning that the evolution of the whole can be specified by the “local factors” as defined here.}

In the current reply, I demonstrate that this claim does not hold.
Even in the product notation, neither the density operator nor its factors simultaneously satisfy~(a) and~(b), because \emph{the so-called `local factors' are not local}.

The central focus of Ref.~\cite{vedral2024locality} concerns a related but more specific issue: identifying the location of phase shifts and other local gates. 
As is well-known, the usual Schrödinger state does not generically permit this identification.
For instance, applying a phase shift (a Pauli~$Z$ gate) to the first qubit of a pair in the state \mbox{$\ket{\Phi^+} = \left(\ket{00} + \ket{11}\right)/{\sqrt 2}$}\, produces \mbox{$({\ket{00} - \ket{11}})/{\sqrt 2}$}, which is the same state that would be obtained if the phase shift were applied to the second qubit.

In the Heisenberg picture of quantum computation, local phase information is encoded in the affected descriptor, and, by virtue of Einsteinian locality, not in other descriptors\footnote{For example, $\ket{\Phi^+}$ can be represented by descriptors $\bol q_1 = (q_{1z}q_{2x}, q_{1x})$ and $\bol q_2 = (q_{2x}, q_{1x}q_{2z})$, assuming that the Heisenberg state is set to $\ket{00}$ (see Ref.~\cite{bedard2021abc} for background on descriptors). Upon performing a $Z$ gate on qubit~1, its descriptor becomes $\bol q'_1 = (-q_{1z}q_{2x}, q_{1x})$, while $\bol q'_2 = \bol q_2$. Had the $Z$ gate been performed on qubit~2, the descriptors would have evolved to $\bar{\bol q}_1 = \bol q_1$ and $\bar{\bol q}_2 = (-q_{2x}, q_{1x}q_{2z})$.}.
Vedral claims that
\lquote{a}{Q2a}{There is also a way of expressing the state in the Schrödinger picture that retains the local knowledge of the phase...}

\noindent I show in~\S\ref{secerror} that this, too, is false. The `locality in the Schrödginger picture' suggested in the title is, therefore, misconceived. 
All quoted passages are from Ref.~\cite{vedral2024locality}.

\section{The Argument}\label{secargument}

First, I present the calculation of Ref.~\cite{vedral2024locality}, which, in Vedral's words,
\lquote{aslocal}{Q3}{
\hspace{-20pt}shows 
that the Schrödinger picture is as Einstein local as the Heisenberg one.}

Consider an initial state $\ket + \ket +$, whose density matrix can be expressed in the so-called \emph{product notation},\vspace{-6pt}
\bes
\vspace{-4pt}
\rho(0) = \frac 14 (I + X_1) (I + X_2) \,,
\ees
where the following notation for Pauli operators is being used:
\beas\vspace{-3pt}
X_1 &=& X \otimes \one \qquad Z_1 = Z \otimes \one \\
X_2 &=&  \one \otimes X \qquad Z_2 = \one \otimes Z  \,.
\eeas
\vspace{-3pt}
Between time~$0$ and~$1$, a controlled-$Z$ gate is applied, so
$$U(1,0) =\operatorname{diag}(1,1,1,-1)= \left(I + Z_1 + Z_2 - Z_1Z_2 \right) / 2 \,.$$ 
This yields \vspace{-2pt}
\be
\vspace{-2pt}
\rho(1) = \frac 14 (I + X_1Z_2) (I + Z_1X_2) \,. \label{eqt1}
\ee
At this stage, we consider applying a phase shift on qubit~$1$, so the evolution between time $1$ and $2a$ is given by $U(2a,1) = Z_1$. 
But before proceeding, it is useful to compare the product notation of~$\rho(1)$, given in Eq.~\eqref{eqt1}, with its tensor-product form,
\bes
\rho(1) = \frac14(I+X\otimes Z+Z\otimes X+Y \otimes Y).
\ees
According to Vedral, 
\lquote{b}{Q4a}{It is this form that tricks us into believing that something non-local is occurring in quantum physics.
Performing the phase operation on the first qubit here leads to the state 
$\rho(2a) = \frac14(I-X\otimes Z+Z\otimes X-Y \otimes Y)$,
and this form just does not tell us which of the two qubits was affected (since [one] could have obtained the same state by a suitable phase kick on the second qubit). In the product notation, on the other hand, the state would become
\bes
\vspace{-2pt}
\rho(2a) = \frac 14 (I - X_1Z_2) (I + Z_1X_2) \,,
\vspace{-2pt}
\ees
which exhibits the minus sign in the state pertaining the first qubit. So, if one looks for a fuller account of what is happening, the product notation is possibly better than the tensor product.}

Although not explicitly discussed in Ref.~\cite{vedral2024locality}, one can verify the effect of applying a phase to the second qubit, i.e., if~$U(2b,1) = Z_2$.
The density operator would become
\bes 
\rho(2b) = \frac 14 (I + X_1Z_2) (I - Z_1X_2) \,,
\ees
with the \emph{second} factor affected. 
As noted in Quote~(Q4a), this is seemingly due to the phase being applied to the \emph{second} qubit.

Vedral confirms the point about~$\rho(2a)$ `exhibi[ting] the minus sign in the state pertaining to the first qubit':
\lquote{qproduct}{Q5a}{...the product notation just introduced for the Schrödinger picture allows us to formally keep track of the dynamics, just like in the Heisenberg picture. Thus, one can also keep track of where local gates have been applied by checking which factor in the product has been affected.}

\noindent As I explain in the next section, this is false.

\section{Why the Product Notation Fails}\label{secerror}

The example given by Verdral, and reexposed in Section~\ref{secargument}, is fine-tuned. I present two modifications of the scenario, both making explicit the fact that the factors in Eq.~\eqref{eqt1} do not pertain to any system.
Thus, the so-called `local factors' are \emph{not} local in any meaningful sense. 

Let us go back to time~$1$, when the state is written as in Eq.~\eqref{eqt1}. 
Instead, consider the effect of applying a local~$X$ gate rather than a~$Z$ gate; \mbox{$U(2c,1) = X_1$}.
In the product notation, the state would become
\bes \label{eq3}
\rho({2c}) = \frac 14 (I + X_1Z_2) (I - Z_1X_2)\,,
\ees
which does \emph{not} affect the first factor in any way, despite it allegedly representing the `state pertaining to the first qubit'.
The action on qubit~$1$ has instead altered the second factor.
Unlike with descriptors for which any action on qubit~1 would only alter its corresponding descriptor, there is no such thing in the product notation as a `state pertaining to the first qubit'.
Conversely, $U(2d,1) = X_2$ would alter the first factor, not the second (and more specifically, it would alter it so as to yield the same state as~$\rho(2a)$, with a minus sign in the first factor).
Moreover, any generic local gate that commutes with neither $X$ nor $Z$ would affect both factors.

To further illustrate the issue, consider the initial state $\ket{\Phi^+} = ({\ket{00} + \ket{11}})/{\sqrt 2}$, whose density matrix in product notation, is given by
\bes
\ketbra{\Phi^+}{\Phi^+} = \frac14 (1 + X_1 X_2)(1 + Z_1Z_2) \,.
\ees
Whether a~$Z$ phase is introduced on qubit~$1$ or qubit~$2$, the first factor is the one affected. Here again, any local gate that commutes with neither $X$ nor $Z$ affects both factors.

Thus, the claim that one can determine the location of a gate by inspecting the factors in the product notation---expressed in Quotes~(Q2a), (Q4a), and (Q5a)---is untenable.
Since these factors do not correspond to individual systems, they fail to satisfy Einsteinian locality.
Consequently, the assertion that the Schrödinger picture is as local as the Heisenberg picture---see Quotes~(Q1) and (Q3)---does not follow from the structure of the product notation.

\section{Internal Contradiction}\label{seccc}

The claims quoted so far are contradicted by other claims, which seemingly acknowledge the failure of the product notation.

For example, the argument in Quote~(Q4a) culminates, in the penultimate sentence, with the (misconceived) assertion that the first factor pertains to the first qubit. 
But the sentence that follows Quote~(Q4a) is a retraction: 
\lquote{criticism-avoiding-claim}{Q4b}{Even here, naturally, there are operations on the second qubit that would lead to the same state...
} 
Indeed, as mentioned in~\S\ref{secerror}, applying an $X$ gate to the second qubit also yields the state~$\rho(2a)$.
In particular, this means that an operation on the \emph{second} qubit alters the \emph{first} factor, namely, what had just been referred to as `the state pertaining the first qubit'.

A similar contradiction appears in Quote~(Q5a), which explicitly asserts that one can track where local gates have been applied by examining which factor in the product has been affected. 
The sentence that follows, however, concedes the point:
\lquote{qprodsuite}{Q5b}{Once more, this only means that, in both pictures, states alone do not contain all the relevant information, and one needs to keep track of the dynamics...} 
Yes. 
In general, in the Schrödinger picture, determining where local gates have been applied requires explicitly tracking the dynamics—that is, the gates themselves.
The passage continues with the correct observation that Heisenberg-picture descriptions do not face this issue:
\lquote{qprodsuite2}{Q5c}{(which, in the Heisenberg picture, is achieved by default by transforming all the algebra of the relevant operators).}

The power of the product notation is also asserted and later retracted in the full version of Quote~\qref{a}, which reads:
\lquote{}{Q2}{There is also a way of expressing the state in the Schrödinger picture that retains the local knowledge of the phase by keeping track of all the operations executed on the system.}

If retaining local phase information requires explicitly tracking all operations, then what is the contribution of the product notation?
After all, there is also a way of using a mathematical constant, for example, $\pi$, to retain the local knowledge of the phase: this is done by leaving $\pi$ aside and independently keeping track of all operations executed on the system.

\bigskip 
In short, insofar as Ref.~\cite{vedral2024locality} asserts the locality of the factors in the product notation, it is incorrect. 
And where it raises the failure of the product notation, it is correct, but self-contradictory. 

Importantly, my critique of the `local factors’ has no impact on another key aspect of the paper—namely, the distinction between Einsteinian and Bell locality, as well as between q-number and c-number-based reality. Vedral argues that Bell’s theorem poses no tension with Einsteinian locality and should instead be understood as a rejection of an underlying c-number-based reality. This interpretation simultaneously undermines the nonlocal, superdeterministic, and retrocausal attempts to explain Bell correlations with c-valued elements. 
I fully endorse this position, as I have shown in Ref.~\cite{bedard2024localarxiv} how a q-number-based reality enables local violations of Bell inequalities.

\subsection*{Acknowledgements}
I am grateful for insightful discussions with Samuel Kuypers, Simone Rijavec, William Schober, Vlatko Vedral and Maria Violaris, which contributed to the development of this work. These discussions took place during my visit to Wolfson College, University of Oxford, which was made possible by Vlatko Vedral’s recommendation and academic support. I especially thank him for facilitating my visit, for valuable exchanges and for including me in the scientific discourse of his research group.

This work was supported in part by the Fonds de recherche du Québec – Nature et technologies, the Swiss National Science Foundation, the Hasler Foundation, and the Mitacs Elevate postdoctoral fellowship in partnership with Bbox Digital.

\bibliographystyle{unsrt}
\bibliography{/Applications/TeX/ref}

\end{document}